\documentstyle[sprocl,psfig]{article}

\def\Journal#1#2#3#4{{#1} {\bf #2}, #3 (#4)}

\def\PRD{{\it Phys. Rev.} D}

\def\lsim{\; \raisebox{-.8ex}{$\stackrel{\textstyle <}{\sim}$}\;}

\def\be{\begin{equation}}
\def\ee{\end{equation}}
\def\bea{\begin{eqnarray}}
\def\eea{\end{eqnarray}}

\def\p{{\mathbf{p}}}

\def\e{\epsilon}

\begin{document}

\title{HIGH ENERGY CONSTRAINTS ON LORENTZ SYMMETRY VIOLATIONS}
\author{S. LIBERATI, T.A. JACOBSON, D. MATTINGLY}
\address{Physics Department, University of Maryland, 
College Park,\\ MD 20742-4111, USA\\
E-mail:liberati@physics.umd.edu, 
jacobson@physics.umd.edu, davemm@physics.umd.edu}
\maketitle

\abstracts{Lorentz violation at high energies might lead to non linear
dispersion relations for the fundamental particles.  We analyze
observational constraints on these without assuming any {\em a priori}
equality between the coefficients determining the amount of Lorentz
violation for different particle species.  We focus on constraints
from three high energy processes involving photons and electrons:
photon decay, photo-production of electron-positron pairs, and vacuum
\v{C}erenkov radiation.  We find that cubic momentum terms in the
dispersion relations are strongly constrained.\\ \\
{\em Talk presented by S.~Liberati at CPT01; the Second Meeting on CPT
and Lorentz Symmetry, Bloomington, Indiana, 15-18 Aug. 2001.}}

\section{Introduction}

There are several reasons to suspect that Lorentz invariance may be 
only a low energy symmetry. This possibility is suggested by 
the divergences of local quantum field theory, as well as by tentative
results in various approaches to quantum gravity. Moreover, the 
unboundedness of the boost parameter makes experimental verification
of exact Lorentz symmetry impossible in principle.

One can study the possibility of Lorentz violation, without a
particular fundamental theory in hand, by considering its
manifestation in dispersion relations for matter. It is natural to
assume that such dispersion relations $E^2(p)$ can be characterized by
an expansion with integral powers of momentum,
\begin{equation}
E^2= p^2+m^2 + \left[A p^2+B p^3/K_0 +C p^4/K_0^2
+O(p^5)\right].
 \label{eq:nldr}
\end{equation}
Here $A,B,C$ are dimensionless coefficients which might be positive as
well as negative and $K_0$ is the ``quantum gravity'' scale (often
identified with the inverse Planck length, $K_0=2\pi/L_{\rm Pl}$).
[Throughout this paper $p$ denotes the absolute value of the
3-momentum vector $\p$, and we use units with the low energy speed of
light in vacuum equal to unity.]

Different approaches to quantum gravity suggest different leading
order Lorentz violating terms. The terms with coefficients $A,B,C$
have mostly been considered so far. Since the $p^2$ term is not
suppressed by the Planck scale, it might be thought to be largest,
however observations severely limit $A$ to be much less than unity
(see e.g.~\cite{CG} and references therein).  The higher order terms
are naturally small, with coefficients of order unity.

Our strategy here is to take a purely phenomenological
stance and consider the constraints that high energy observations  
impose on dispersion relations of the form
\begin{equation}
E_{a}^2 \approx  p_{a}^2+m^{2}_{a}  +
 \eta_{a} {p_{a}^n}/{K_0^{n-2}},
 \label{eq:pdr}
\end{equation}
where $a$ labels different fields and $n\ge3$. In the absence of a
fundamental theory one has no reason to expect any particular
relation between the coefficients $\eta_{a}$ for different particles,
except perhaps that they should all be of the same order of magnitude.

Dispersion relations of the kind (\ref{eq:pdr}) produce kinematic
constraints from energy-momentum conservation that differ from the
usual Lorentz invariant case.  As a result reactions can take place
that are normally forbidden, and thresholds for reactions are
modified.  Observational consequences may seem out of reach because of
the Planck scale suppression of the Lorentz violating terms (assuming
that, as generally expected, $K_{0}$ is of order the Planck scale). 
However this is not so.  One can expect deviations from standard
kinematics when the the last two terms of (\ref{eq:pdr}) are of
comparable magnitude.\footnote{In general of course one
must look at the specific reaction in order to estimate the energies
at which deviations from standard behavior can be expected.  For
example in the case of the reaction $\gamma\gamma\rightarrow
e^{+}e^{-}$ it will be the electron mass that sets the scale.}
Assuming $\eta$ is of order unity this yields the condition $p_{\rm
dev}\sim (m^2 K^{n-2}_{0})^{1/n}$, which is $\sim
(m/m_e)^{2/3}\times 10$ TeV for $n=3$ and $\sim (m/m_e)^{1/2}\times
10^4$ TeV for $n=4$.  Although these energies are currently not
achievable in particle accelerators (except in the case of the massive
neutrinos which however are weakly coupled) they are in the range of
current astrophysical observations.

In fact, it has been suggested by several
authors~\cite{CG,ACP,Mestres,Bertolami,Kifune,Kluzniak,Aloisio} (see
also \cite{Sigl} and references therein) that we may already be
observing deviations from Lorentz invariance via the existence of two
puzzles in modern astrophysics: the missing GZK cut-off on
cosmic ray protons with ultra high energy greater than $7\times
10^{19}$ eV, and the apparent overabundance of gamma rays above 10 TeV
from the BL Lac system Mkr 501~\cite{Protheroe}.  Here we shall mostly
not consider the constraints imposed by asking Lorentz violation to
explain these puzzles.  Instead we restrict our attention to
constraints imposed by consistency with known phenomena (or lack
thereof). (See also~\cite{CG,Mestres,Kifune} for a similar
discussion.)


\section{Observational constraints}

To find the strongest observational constraints without assuming {\it
a priori} relations between the coefficients $\eta_a$ of
(\ref{eq:pdr}) we focus here on processes involving just two
fundamental particles, photons and electrons.  We also restrict to the
case $n=3$, since this should be most tightly constrained. If it can
be ruled out, one can then move on to the $n=4$ case.

The modified dispersion relations for photons and electrons in general
allow two normally forbidden interactions: photon decay,
$\gamma\rightarrow e^+e^-$, and vacuum \v{C}erenkov radiation,
$e^-\rightarrow e^-\gamma$.  If allowed these processes happen very
rapidly. In addition the threshold for photon annihilation,
$\gamma\gamma\rightarrow e^+e^-$, is shifted.  We consider constraints
that follow from three observations: (i) Gamma rays up to $\sim 50$
TeV of cosmological origin arrive on earth~\cite{Tanimori}, so photon
decay does not occur up to this energy.  (ii) Electrons of energy
$\sim 100$ TeV are believed to produce observed X-ray synchrotron
radiation coming from supernova remnants.  Assuming these electrons
are actually present, vacuum \v{C}erenkov radiation must not occur up
to that energy.  (iii) Cosmic gamma rays below 10 TeV are believed to
be absorbed in a manner consistent with photon annihilation off the IR
background with the standard threshold.  Observation (iii) is not
model independent, so the corresponding constraint is tentative and
subject to future verification.

To derive the observational constraints one needs to determine the
threshold for each the process, i.e. the lowest energy for which the
process occurs. The details concerning our determination of the
thresholds are reported elsewhere~\cite{JLM01}.  Assuming monotonicity
of $E(p)$ (for the relevant momenta $p\ll K_0$) we have shown that all
thresholds for processes with two particle final states occur when the
final momenta are parallel. Moreover for two particle initial states
the incoming momenta are antiparallel. This geometry has been assumed
in previous works but to our knowledge it was not shown to be necessary.
(In fact it is not necessary if $E(p)$ is not monotonic.)

To eliminate the subscript $a$ we introduce $\xi\, :=\eta_\gamma$ and
$\eta\, := \eta_e$.  The constraints will restrict the allowed region
of the $\eta$--$\xi$ plane.  For the rest of the paper we assume $K_0$
is the Planck energy and we use units with $K_0=1$.

\subsection{Photon decay}

Photon decay is allowed only above a broken line in the $\eta$--$\xi$
plane given by $\xi=\eta/2$ in the quadrant $\xi,\eta>0$ and by
$\xi=\eta$ in the quadrant $\xi,\eta<0$. Above this line, the
threshold is given by
\begin{eqnarray}
k_{\rm th}&=& \displaystyle{\left( \frac{8 m^2}{2\xi-\eta} \right)^{1/3} 
\quad \quad \:\:\,\mbox{for $\xi\geq0$}},\label{th1}
\\
k_{\rm th}&=&
\displaystyle{\left( \frac{-8 m^2\eta}{(\xi-\eta)^2} \right)^{1/3}} 
\qquad \mbox{for $\eta<\xi<0$.}\label{th2}
\end{eqnarray}
The first relation (\ref{th1}) arises when the electron and positron
momenta are equal at threshold. In standard Lorentz invariant
kinematics such ``equipartition" of momentum always holds at threshold
for pair production. In all previous work on Lorentz violating
dispersion it has been assumed to hold. Surprisingly, however, in the
present case the threshold may occur with an asymmetric distribution
of momentum.  The second relation (\ref{th2}) applies in those cases.
The constraint we impose is that the threshold is above 50 TeV, the
energy of highest observed gamma rays from the Crab nebula~\cite{Tanimori}.

\subsection{Photon annihilation}
\label{photann}
The threshold relations for a gamma ray to annihilate with an IR
background photon of energy $\e$ take approximately the same
form as for photon decay, with the replacement $\xi\rightarrow\xi^{'}$,
where $\xi^{'}\equiv \xi+4\e/k_{\rm th}^2$. (Here we have used
the fact that $\e$ is much smaller than any other scale in the
problem.) The two different relations arise for the same reason as in
the case of photon decay, and they correspond respectively to cubic
and quartic polynomial equations for $k_{\rm th}$. 

For the observational consequences it is important to 
recognize that the threshold shifts are much more
significant at higher energies than at lower energies. 
To exhibit this dependence, it is simplest to fix a 
gamma ray energy $k$ and to solve for the corresponding 
soft photon threshold energy $\e_{\rm th}$. Taking the
ratio with the usual threshold $\e_{{\rm th},0}$, 
we find a dependence on $k$ at least as strong as $k^{3/2}$.
Introducing $k_{10}:=k/(10\,\mbox{TeV})$, we have
\begin{eqnarray}
\frac{\e_{\rm th}}{\e_{{\rm th},0}}&=& 
1+0.05\,(\eta-2\xi)\,k^{3}_{10}
\quad \qquad \qquad  \:\:\:\,\mbox{for $\xi'\geq0$},\label{eth1}\\
 \frac{\e_{\rm th}}{\e_{{\rm th},0}}&=&
0.1\,(\eta-\xi)\,k^{3}_{10}+\sqrt{-0.2\,\eta\, k^{3}_{10}}
 \qquad \mbox{for $\eta<\xi'<0$.}\label{eth2}
\end{eqnarray}

High energy TeV gamma rays from the blazars Markarian 421 and
Markarian 501 have been detected out to 17 TeV and 24 TeV
respectively.  Although the sources are not well understood, and the
intergalactic IR background is also not fully known, detailed modeling
shows that the data are consistent with some absorption by photon
annihilation off the IR background (see
e.g.~\cite{Mrk421,JStecker01,Aharo2} and references therein). 
However, while the inferred source spectrum for Mrk 501 is consistent 
with
expectations for energies less than around 10 TeV, above this energy
there are far more photons than expected according to some IR
background models~\cite{Protheroe,Aharo2}.
 
The uncertainty in the blazar and IR background models currently
precludes sharp constraints from photon annihilation.  Instead, we
just determine the range of parameters $\xi,\eta$ for which the
threshold $k_{\rm th}$ lies between 10 TeV and 20 TeV for an IR photon
with which a 10 TeV photon would normally be at threshold.  We choose
this range since (i) lowering the threshold would make the
overabundance problem worse, and (ii) raising this threshold by a
factor of two might explain the potential overabundance of photons
over 10 TeV~\cite{ACP}.  If the overabundance puzzle exists and is to
be resolved in this way, the effect of the threshold shifts must be
enhanced for photons above 10 TeV relative to those below 10 TeV
(since the latter seem to be well accounted for).  This enhancement
could arise partly from the shape of the IR background spectrum, but
the shift itself is also enhanced, as seen in the energy dependence of
equations (\ref{eth1},\ref{eth2}). The
hypothesis~\cite{ACP,Kifune,Protheroe} that the potential
overabundance is due to Lorentz violating dispersion with $n=3$
therefore appears to be consistent with current observations.

\subsection{Vacuum \v{C}erenkov radiation}

An electron can emit \v{C}erenkov radiation in the vacuum if $\eta>0$
or if $\eta<0$ and $\xi<\eta$.  In this case there are also two
threshold relations, depending on whether the threshold occurs with
emission of a zero-energy photon or with emission of a finite energy
photon.  These two cases correspond to the two following relations,
respectively:
\begin{eqnarray}
\quad p_{\rm th} &=& 
\displaystyle{\left(\frac{m^2}{2\,\eta}\right)^{1/3}}
\qquad \qquad \qquad \:\:\: \mbox{for $\eta > 0$ and $\xi\geq-3\eta$},
\label{Cerenkov1}\\
\quad p_{\rm th} &=& 
\displaystyle{\left(-\frac{4\,m^2\left(\xi+\eta\right)}
{\left(\xi-\eta\right)^2}\right)^{1/3}}
\qquad 
\mbox{for $\xi<-3\eta<0$  or $\xi<\eta \leq 0$.}
\label{Cerenkov2}
\end{eqnarray}
The vacuum \v{C}erenkov process is extremely efficient, leading to an
energy loss rate that goes like $E^2$ well above threshold.  
Thus any electron known to
propagate must lie below the threshold.  
Electrons of energy $\sim 100$ TeV are believed to produce observed 
X-ray synchrotron radiation coming from supernova remnants.
Thus for example in the region of the parameter plane
where (\ref{Cerenkov1}) holds we obtain the constraint
$\eta<m^2/2p_{\rm th}^3\sim 10^{-3}$.

\section{Combined constraints}

\begin{figure}[ht]
\begin{center}
\centerline{\psfig{figure=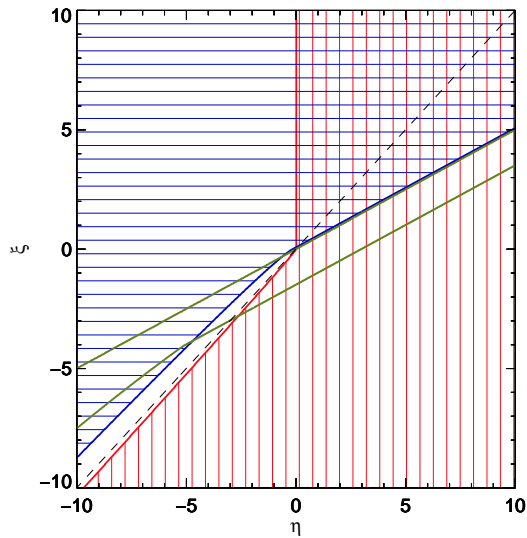,height=3in}}
\caption{Combined constraints on the photon and electron parameters
for the case $n=3$. The regions excluded by the photon decay and
\^{C}erenkov constraints are lined horizontally in blue and vertically
in red respectively. The photon annihilation threshold limits fall
between the two diagonal green lines. The dashed line is
$\xi=\eta$.}~\label{fig:all}
\end{center}
\end{figure}

Putting together all the constraints and potential constraints we
obtain the allowed region in the $\eta$--$\xi$ plane (see
Figure~\ref{fig:all}).  The photon decay and \v{C}erenkov constraints
exclude the horizontally and vertically shaded regions,
respectively. The allowed region lies in the lower left quadrant,
except for an exceedingly small sliver near the origin with
$0<\eta\lsim 10^{-3}$ and a small triangular region
($-0.16\lsim\eta<0$, $0<\xi\lsim0.08$) in the upper left quadrant.
The range of the photon annihilation threshold discussed in subsection
\ref{photann} falls between the two roughly parallel diagonal
lines. This intersects the otherwise allowed region in a finite,
narrow wedge where $\xi$ and $\eta$ are negative and of order unity
(apart from a minuscule invisible region near the origin with
$\eta>0$).
If future observations of the blazar fluxes and the IR background
confirm agreement with standard Lorentz invariant kinematics, the region
allowed by the photon annihilation constraint will be squeezed toward
the upper line ($k_{\rm th}\approx k_{\rm s}$).  If the overabundance
of gamma rays from Mrk 501 is indeed due to Lorentz violation of the
sort we are considering, the region will be squeezed toward the lower
line ($k_{\rm th}\approx 2k_{\rm s}$) or may even shift toward lower
values of $\xi$ and $\eta$ if a yet larger threshold shift is
indicated.

\section{Conclusions}

We have seen that a conservative interpretation of observations puts
strong constraints on the possibility of Planck scale cubic
modifications to the electron and photon dispersion relations.  The
allowed region includes $\xi=\eta=-1$, which has been a focus of
previous work~\cite{ACP,Kifune,Kluzniak}.  The negative quadrant has
most of the allowed parameter range. It is interesting to note that in
this quadrant all group velocities are less than the low energy speed
of light.

To completely rule out the cubic case would require new observations.
Finding higher energy electrons would not help much, while finding
higher energy undecayed photons would squeeze the allowed region onto
the line $\xi=\eta$.  To further shrink the allowed segment of this
line would require observations allowing the usual threshold for
photon annihilation to be confirmed to higher precision.  Perhaps
other processes could be used as well. If {\it a priori} relations
among the coefficients in the dispersion relations for different
particles are hypothesized, stronger constraints can of course be 
obtained.\\

\noindent{\em This research was supported in part by NSF grant PHY-9800967.}
 
\section*{References}


\begin{thebibliography}{99}
\bibitem{CG}
S.~Coleman and S.~L.~Glashow,
\Journal{\PRD}{59}{116008}{1999}.

\bibitem{ACP}
G.~Amelino-Camelia and T.~Piran,
{\em Phys.\ Rev.\ }D{\bf 64}, 036005 (2001).

\bibitem{Mestres}
L.~Gonzalez-Mestres,
{\em ``Gamma and cosmic ray tests of special relativity''},
astro-ph/0011181 and references therein.

\bibitem{Bertolami}
O.~Bertolami,
{\em ``Ultra-high energy cosmic rays and symmetries of spacetime''},
astro-ph/0012462.


\bibitem{Kifune}
T.~Kifune,
{\em Astrophys.\ J.\ } {\bf 518}, L21 (1999).

\bibitem{Kluzniak}
W.~Kluzniak,
{\em Astropart.\ Phys.\ }{\bf 11}, 117 (1999).

\bibitem{Aloisio}
R.~Aloisio, {\em et al.},
{\em Phys.\ Rev.\ }D{\bf 62}, 053010 (2000).

\bibitem{Sigl}
G.~Sigl,
{\em Lect.\ Notes Phys.\ }{\bf 556}, 259 (2000).

\bibitem{Protheroe}
R.~J.~Protheroe and H.~Meyer,
{\em Phys.\ Lett.\ }B{\bf 493}, 1 (2000).

\bibitem{Tanimori}
T.~Tanimori {\em et al.}, {\em ApJ} {\bf 492}, L33 (1998).

%
\bibitem{JLM01}
T.~Jacobson, S.~Liberati, and D.~Mattingly,
{\em In Preparation}.

\bibitem{Mrk421}
F.~Krennrich {\it et al.},
{\em ``Cutoff in the TeV energy spectrum of Markarian 421 during strong flares  
in 2001''},
astro-ph/0107113.

\bibitem{JStecker01}
O.~C.~de Jager and F.~W.~Stecker,
{\em ``Extragalactic gamma-ray absorption and the intrinsic spectrum of Mkn 
501 during the 1977 flare,''}
astro-ph/0107103.

\bibitem{Aharo2}
F.~A.~Aharonian, A.~N.~Timokhin and A.~V.~Plyasheshnikov,
{\em ``On the origin of highest energy gamma-rays from Mkn-501,''}
astro-ph/0108419.

\end{thebibliography}
\end{document}